\documentclass[letterpaper,preprint,rsi,superscriptaddress,floatfix]{revtex4}
\pdfoutput=1
\usepackage[german,swedish,english]{babel}
\usepackage[dvips]{graphicx}
\usepackage{amssymb}
\usepackage{amsmath}
\usepackage{color}

\begin{document}

\date{\today}

\title{Time-resolved one-dimensional detection of x-ray scattering in pulsed magnetic fields}
\author{Zahirul Islam}
\author{Jacob P. C. Ruff}
\affiliation{X-Ray Science Division, Advanced Photon Source, Argonne National Laboratory, 9700 S. Cass Ave., Argonne IL 60439\\}
\author{Kate A. Ross}
\affiliation{Department of Physics and Astronomy, McMaster University, Hamilton, ON, L8S 4M1, Canada\\}
\author{Hiroyuki Nojiri}
\affiliation{Institute for Materials Research, Tohoku University, Sendai, Japan\\}
\author{Bruce D. Gaulin}
\affiliation{Department of Physics and Astronomy, McMaster University, Hamilton, ON, L8S 4M1, Canada\\}
\affiliation{Brockhouse Institute for Materials Research, McMaster University, Hamilton, ON, L8S 4C6, Canada\\}
\affiliation{Canadian Institute for Advanced Research, 180 Dundas Street West, Toronto, ON, M5G 1Z8, Canada}

\begin{abstract}
We have developed an application of a one-dimensional micro-strip detector for capturing x-ray diffraction data in pulsed magnetic fields. This detector consists of a large array of 50 $\mu$m-wide Si strips with a full-frame read out at 20 kHz. Its use substantially improves data-collection efficiency and quality as compared to point detectors, because diffraction signals are recorded along an arc in reciprocal space in a time-resolved manner. By synchronizing with pulsed fields, the entire field dependence of a two-dimensional swath of reciprocal space may be determined using a small number of field pulses.
\end{abstract}

\maketitle

The use of high field pulsed magnets at synchrotron x-ray radiation facilities \cite{minicoil1,minicoil2,xdiff1,pow1,PvdL,apspfm} has emerged as a viable approach for studying materials in magnetic fields well beyond 15 T \cite{hifield1,hifield2,nrc,cmmp2010,xtreme,jpn-100T}. Several pulsed magnets in use generate peak fields in the range of 30-40 T with a full width of a few milliseconds for a half-sine pulse shape. For measurements requiring maximum fields, as much as $\sim$20 minutes of waiting in between successive shots may be required for the magnet to cool down to operating temperature, and therefore it is critical to extract as much information as possible from each pulse. Due to high brightness of the third-generation synchrotron sources, these pulsed magnets have been successfully used in numerous studies \cite{pow1,pow2, xdiff1,xdiff2,xdiff3,xdiff4,xdiff5,xmcd,xmcd2,xres,val}; nevertheless, increases in measurement efficiency are highly desirable. In this article we describe a novel use of a fast one-dimensional $\mu$-strip detector for diffraction measurements. Such detectors are typically used for spectroscopic studies using energy dispersive optics (see Ref.\ \onlinecite{strip1,strip2}) and powder diffraction\cite{strip3}. However, the use of strip detectors has substantially improved data collection efficiency and quality by enabling one to record diffraction intensity along arcs in reciprocal space in time-resolved fashion in order to collect field dependence using a small number of field pulses.

\begin{figure}[ht]
\includegraphics[width=1.0\textwidth]{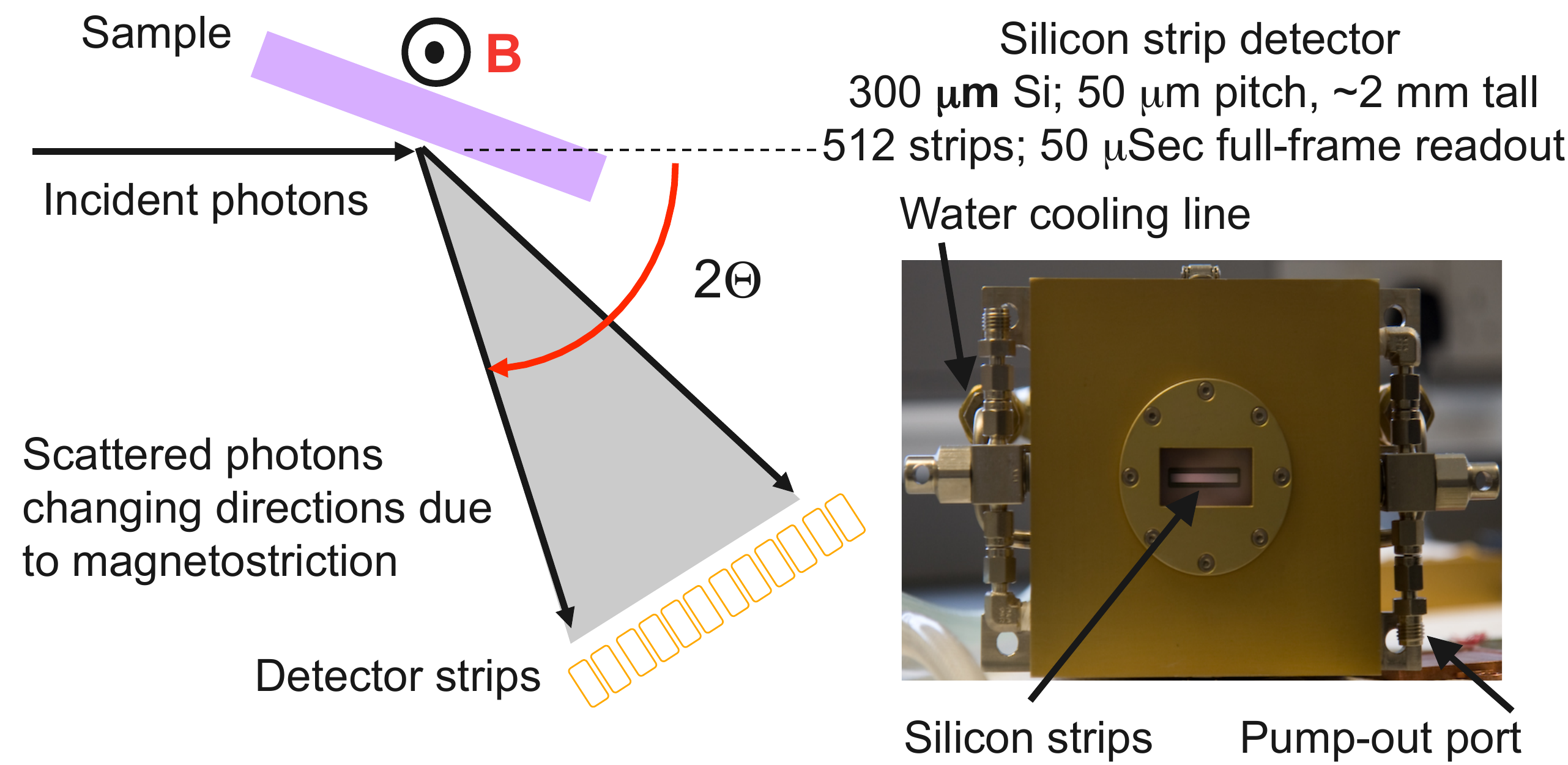}
\caption{Left: Time-resolved data collection scheme using a $\mu$strip detector. Right: Front view of the strip detector with the Be window removed in order to display the active Si region. During the field pulse the detector is exposed for 20 $\mu$Sec. and the full frame is read out every 50 $\mu$Sec (20 kHz) capturing the angular shifts of Bragg peak in $2\theta$ due to magnitostriction.}
\label{strip}
\end{figure}

Our detector is a prototype ULTRA model manufactured by Quantum Detectors (Fig.\ \ref{strip}). The magnet itself is described in detail elsewhere \cite{apspfm}. The detector consists of 512 Si strips each of which is 50$\mu$m wide, 300$\mu$m thick, and $\sim$2 mm in height. The active area of the detector head is kept under vacuum, with a Be window for x-ray access. Using a combination of water and thermo-electric cooling the detector head is kept at -39$^\circ$ C. Each strip provides an integrated photon count with 16 bits of dynamic range. In order to collect time-resolved diffraction intensity accurately, the strip detector, magnetic-field pulse generation, and data acquisition instruments need to be well synchronized. The entire synchronization process is carried out via hardware triggers (TTL, transistor-transistor logic pulses) produced by a digital delay generator (SRS DG535). The strip detector  is configured to integrate photon counts for 20 $\mu$Sec. and triggered using a TTL pulse pattern generator (Model 81110A from Agilent Technologies) in order to store full-frame data at 20 kHz. Typically a train of 40  TTL pulses is sent to the strip detector in order to record full-frame data for 2 ms spanning the total duration of the field pulse of $\sim 1$ ms \cite{apspfm}. The data are recorded over a Gigabit Ethernet connection to a PC. Pulsed current profiles from a high-precision current monitor as well as pulse start and end signals from the capacitor bank are captured by a 10MHz digital storage oscilloscope (DSO; LDS model Sigma 30). In the final step, all the data from PC and DSO are transferred into the beamline workstation for storage and analysis. The detection scheme was tested for single-crystal diffraction studies on X-ray Operations and Research (XOR) 4-ID-D and 6-ID-B beamlines. A monochromatic (7.114 keV and 10.885 keV on 4-ID-D, and 16.2 keV on 6-ID-B) beam of x-rays was selected using a double-bounce Si(111) monochromator. Two mirrors were utilized to focus and reject harmonics in the beam. A CHESS-type ion chamber filled with helium gas was used to monitor incident flux.

The basic idea of using a strip detector is shown in Fig.\ \ref{strip}. The strip detector is mounted on an XY-translation stage attached to the 2$\theta$ arm of a Huber diffractometer so that all the strips lie in the scattering plane. The linear array of the strips then forms an arc in the reciprocal lattice with varying 2$\theta$ angles. The distance of the strips to the sample was $\sim$1 m giving an angular resolution of $\Delta (2\theta) < 0.003^\circ$. The goal then is to record intensity of all strips in time bins of width $\delta t$ at a rate of 20 kHz of a single-crystal Bragg peak that illuminates several neighboring strips. Since the crystal is kept fixed during field pulse, several sets of time-resolved data are collected for different angular orientations of the crystal with respect to the incident x-ray beam in order to collect integrated intensity and measure field dependence of a 2D swath of reciprocal space.

\begin{figure}[ht]
\includegraphics[width=0.7\textwidth]{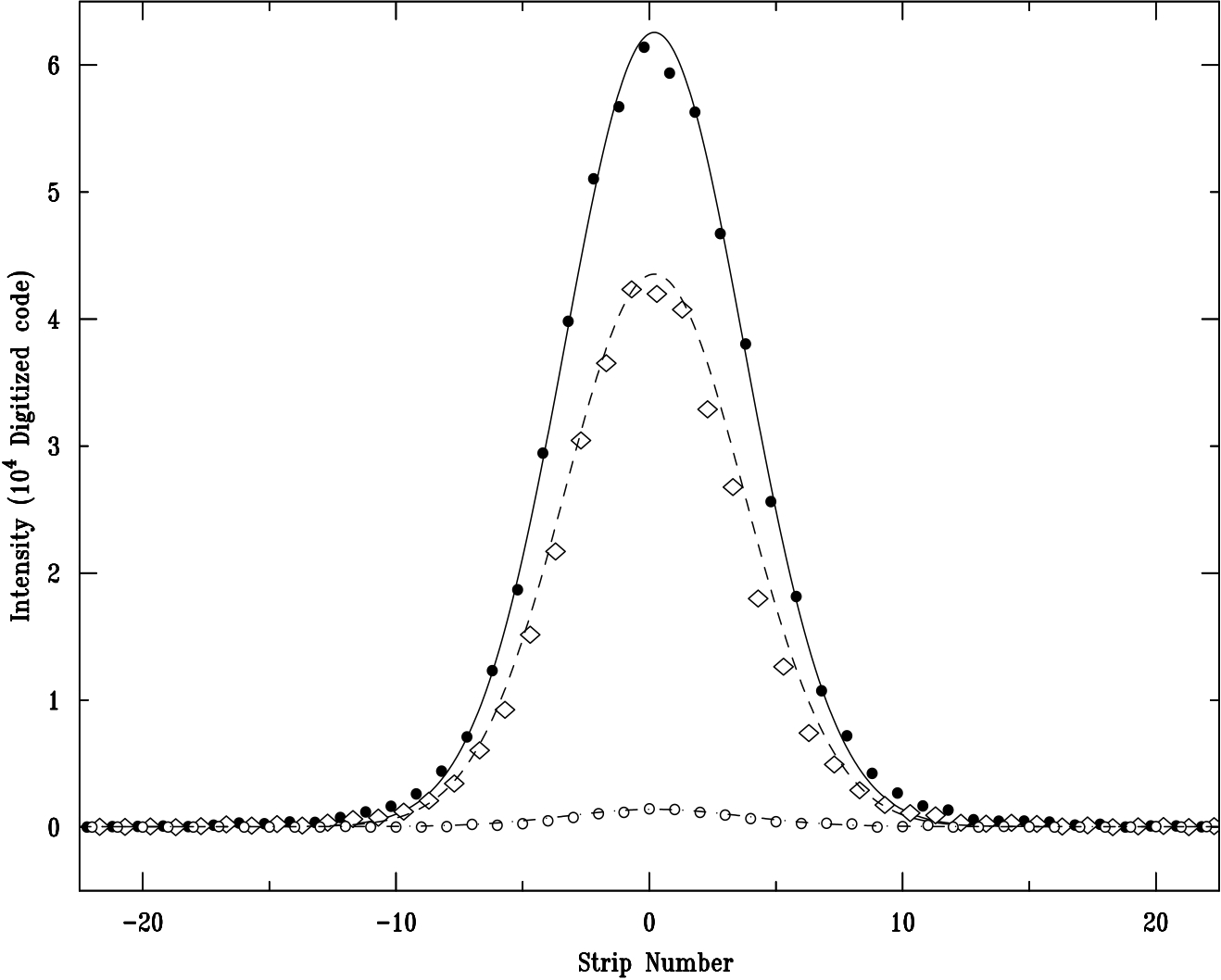}
\caption{Line shape of the incident beam, with three values of photon flux differing by more than two orders of magnitude. The integration time was 20 $\mu$S. Different symbols correspond to different magnitudes of incident flux.}
\label{lnshape}
\end{figure}

In commissioning the strip detector, we have checked the linearity and dynamic range of the detector. Since the pulsed field duration is $\sim$1ms, we need to record diffracted intensity in $\delta t\leq 20$ $\mu$S, or smaller time bins in order to obtain a good field resolution. Therefore, it is important to determine the linearity and dynamic range per strip for 20 $\mu$S integration. The tests were performed by illuminating the detector by the incident x-ray beam. The beam size as measured by scanning a blade (placed before the detector) across the beam was $\sim100$ $\mu$m X 300 $\mu$m in the vertical and horizontal directions, respectively. The total photon flux was varied by  detuning the undulator and was monitored by a He-gas filled ion chamber. Fig.\ \ref{lnshape} shows the intensity profile of the incident beam as it impinged on the strip detector. The beam profile can be well modeled by a Gaussian with a full width (in the horizontal plane) of $\sim350 \mu$m for flux varying over two orders of magnitude.

\begin{figure}[ht]
\includegraphics[width=0.8\textwidth]{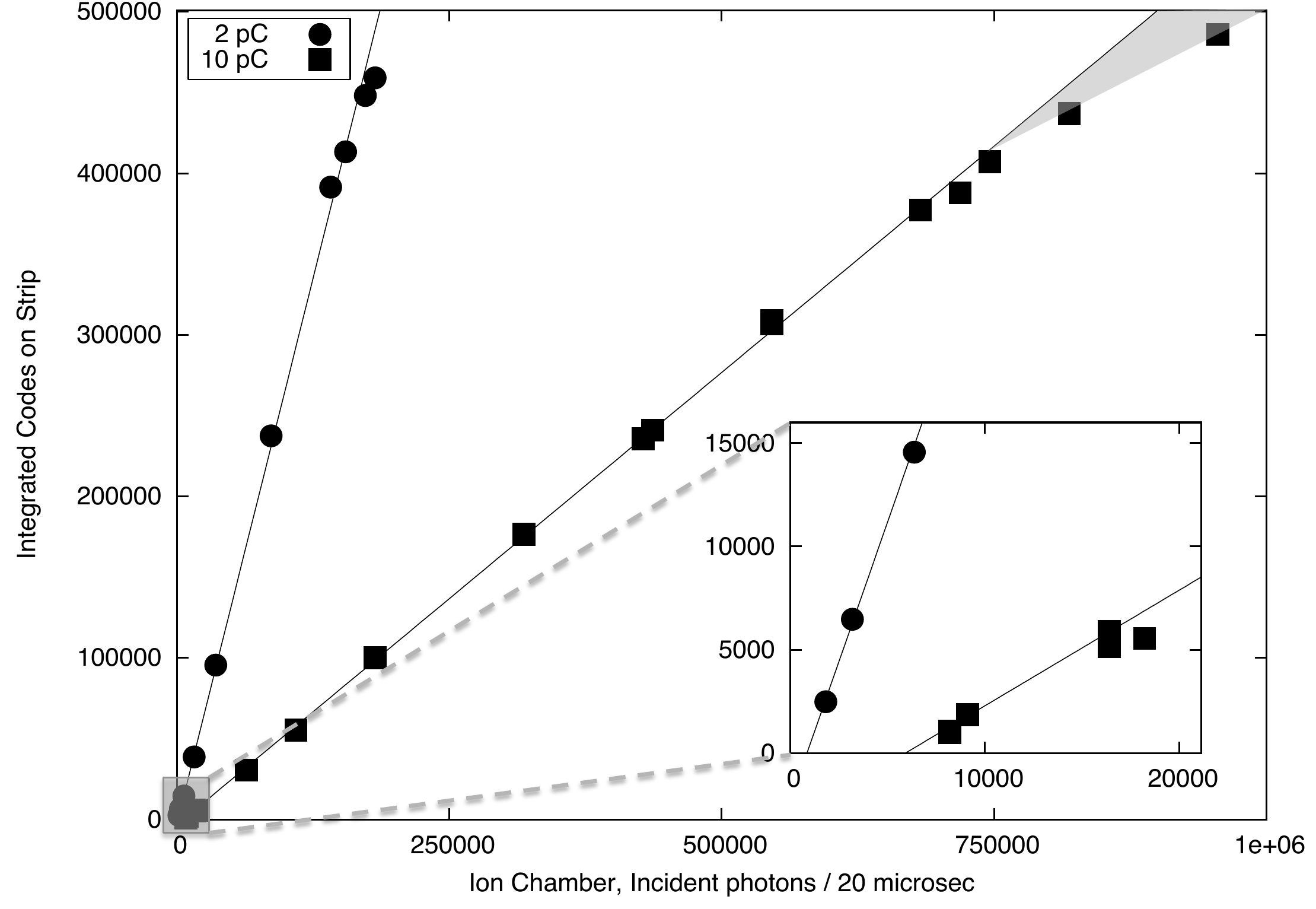}
\caption{Integrated codes (proportional to photon counts) as a function of increasing photon flux collected using 2 pC and 10 pC settings.  Deviations from linearity at very high photon flux for 10 pC setting are indicated by a shaded triangle. The inset shows the data near the origin on an expanded scale. The threshold for peak detection is lower for the 2 pC setting.}
\label{linearity}
\end{figure}

Fig.\ \ref{linearity} shows the integrated intensity derived from the area of the beam profile as a function of photon flux impinging on the detector. The detector was exposed for 20 $\mu$S and read out at 20 kHz, a scheme used during pulsed-field studies alluded to above. The flux was increased to determine the maximum count rate needed to saturate an individual strip. During the measurements total flux was always kept below this value such that individual pixels at beam center were not saturated. The detector can be used with two sets of capacitors corresponding to 2 pC and 10 pC of charge. These two settings allow one to operate over a large range of incident photon flux. In each case, the detector response is linear as the incident flux is varied by two orders of magnitude. With the full 10 pC setting the detector is almost  linear with incident flux exceeding $10^{10}$ photons/sec. Note that there is a small offset indicating a minimum threshold of incident photon flux, below which a Bragg peak is not discernible in a single 20 $\mu$Sec exposure. Of particular interest is the 2 pC setting which has a smaller threshold for peak detection. This latter setting is suitable for measuring Bragg peaks that are less intense, or may become weaker with magnetic fields. 

\begin{figure}[ht]
\includegraphics[width=1.0\textwidth]{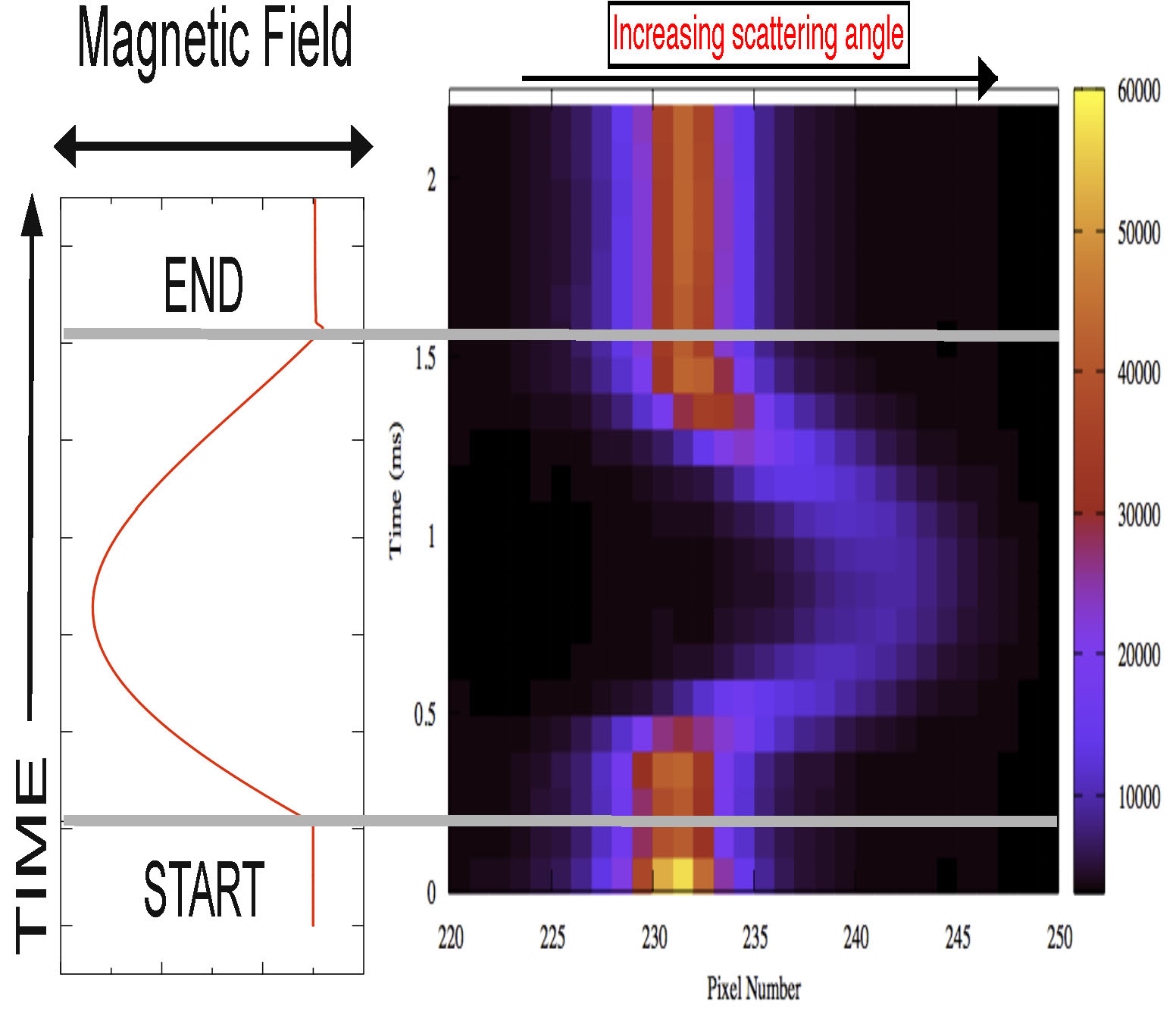}
\caption{Left: Typical pulsed field profile. Right: 2D time-angle map of diffracted intensity of (8, 0, 0) peak captured using a 1D strip detector for a peak field of 20 Tesla. A pair of horizontal grey lines indicate pulse start and end times.}
\label{strip-dat}
\end{figure}

For tests of the detection scheme we measured magnetostriction  (MS) effects of a single-crystal Tb$_2$Ti$_2$O$_7$ sample \cite{jetp1,jetp2,ruff1,ruff2}. The split-pair pulsed magnet and a single-crystal sample were cooled using a dual-cryostat system \cite{apspfm}. We recorded Bragg peak intensity in the strip detector as a function of time before, during, and after the field-pulse generation. Since $\sim 1$ ms is the dwell time of the pulsed field we measured scattered photons as a function of time for $\sim 2$ ms. A $\sim 20$ $ \mu$Sec integration at 20 kHz full frame read out provided field dependence of a Bragg peak using a single pulse. If the lattice parameters contract (expand) due to MS effects induced by pulsed magnetic fields then the Bragg peak would shift to higher (lower) angles. Accordingly the peak center would move to different strips as a function of time, directly yielding MS shifts as a function of field. Fig.\ \ref{strip-dat} shows how the intensity of a Bragg peak varies as a function of field as monitored along an arc in the reciprocal space during a single pulse. Intensity of the Bragg peak observed at the zero-field reciprocal-latice point is fully suppressed as the lattice parameter shrinks with time due to pulsed fields. As a result the intensity is observed at strips corresponding to higher $2\theta$ angles. From the size of individual strip width and their known distance from the sample, their angular shifts can be obtained. For example for the data shown, (8, 0, 0) Bragg peak shifts as much as $\sim0.045^\circ$ from its zero-field position. Note that measurements of the full integrated intensity required a set of time-resolved data to be collected as the crystal is ``rocked'' through a Bragg peak as in point-by-point method.

We have implemented a fast one-dimensional strip detector method, to collect time-resolved diffraction intensity along an arc in the reciprocal space. The use of this strip detector goes a long way in increasing data collection efficiency over a large dynamic range which is critical, given the low duty cycle of pulsed fields. The use of the strip detector for diffraction was demonstrated by studying MS effects in a geometrically frustrated magnet, Tb$_2$Ti$_2$O$_7$. The application of strip detector in this work makes it clear that a major progress in broader pulsed-magnet use in x-ray studies is possible with the advent of fast area detectors. The use of such fast detectors may usher in novel studies of field-induced phases in the near future. 

We appreciate R. Goldsbrough (Quantum Detectors) and A. Micelli (APS) for technical assistance with the strip detector and its control software. Use of the APS is supported by the DOE, Office of Science, under Contract No. DE-AC02-06CH11357. A part of the is supported by International Collaboration Center at the Institute for Materials Research (ICC-IMR) at Tohoku University. HN acknowledges KAKENHI No.  23224009 from MEXT. JPCR, BDG, and KR acknowledge the support of NSERC of Canada.

\end{document}